# WIKI THANKS: CULTURAL DIFFERENCES IN THANKS NETWORK OF DIFFERENT-LANGUAGE WIKIPEDIAS

Keiichi Nemoto and Ken-ichi Okada

Keio University
3-14-1 Hiyoshi, Kohoku-ku Yokohama, Kanagawa, 223-8522, Japan
e-mail: nemoto@a7.keio.jp

## ABSTRACT

Wikipedia introduced a new social function "wiki-thanks". "Wiki-thanks" enable editors to send thanks to other editors' contributions. In this paper, we aim to investigate this new social tool from different cultural perspectives. To achieve this goal, we analyze "wiki-thanks" log events and compared the English, German, Spanish, Chinese, Japanese, Korean, and Finish language Wikipedias.

## INTRODUCTION

In the process of Wikipedia growth, there are a lot of aspects of collaboration in different stage of article building and decision-making process. Scholars have investigated the interaction of editors and the effect of production performance (Kitter 2008, Nemoto 2011). According to the 2011 survey of Wikipedia editors[1], most people agree that types of messages from other editors affect the frequency of edits. Compliment increase frequency and contempt from experienced editors decrease frequency of edits.

From the cross-cultural perspective, Wikipedia exists in 288 languages (February 15, 2015), among them are languages like Finish, Korean, and Japanese which are not shared by other countries. Therefore Wikipedia offers a kind of microscope to analyze how people in these local cultures work together.

In this paper, we aim to identify cultural differences in online collaboration settings and the difference of motivational factors among different language Wikipedias. In order to investigate the difference, we focus on feedback systems deployed on the Wikipedia called "wiki-thanks".

## WIKI-THANKS

Wikipedia deployed Thanks Notifications[2] on May 30, 2013. This feature is easier way to send positive expression - "thanks"- to other Wikipedians than the "Wiki-love"[3] introduced on October 2011. This new feature allows registered editors to send thanks to other editors using a link corresponding to each edit history in the "View history tab" such as a FaceBook Like button.

## Data

All Wiki-thanks logs are stored and can be retrieved from MediaWiki API service. The log data contains a set of items such as when the log event created, who sent "thanks" to whom. Wiki-thanks network can be created from these log data. Table 1 shows the number of unique users who sent or received wiki-thanks and the total number of events up to November 2014. Wiki-thanks ratio per edits is the highest in German Wikipedia and lowest in Korean Wikipedia.

*Table1: Total number of wiki-thanks and wiki-thanks adoption ration on July 2014*

| | Launch month | Total wiki-thanks | Edits per Month (July 2014) | # of wiki-thanks(July 2014) (per edits) | Active Editors (July 2014) | # of unique sender and receipient(July 2014) (per active editors) |
|---|---|---|---|---|---|---|
| ja | Sep-2013 | 19972 | 259375 | 2000 (0.771%) | 3915 | 1051 (26.85%) |
| zh | Oct-2013 | 11277 | 218559 | 1288 (0.589%) | 2378 | 587 (24.68%) |
| en | May-2013 | 302946 | 2850217 | 22573 (0.792%) | 31819 | 9540 (29.98%) |
| de | Nov-2013 | 85595 | 539147 | 7529 (1.396%) | 6001 | 2982 (49.69%) |
| ko | Sep-2013 | 2011 | 99671 | 156 (0.157%) | 810 | 111 (13.70%) |
| fi | Oct-2013 | 3215 | 49938 | 265 (0.531%) | 472 | 161 (34.11%) |
| es | Sep-2013 | 29313 | 509967 | 2688 (0.527%) | 4142 | 1148 (27.72%) |

Following the English (en) Wikipedia, thanks feature was introduced on other language Wikipedia. From the log analysis of wiki-thanks event, it was launched on September 2013 in the Japanese (ja), Spanish (es), and Korean (ko) Wikipedia, October 2013 in the Chinese and Finish (fi) Wikipedia, and November 2013 in the German (de) Wikipedia.

After the launch, the number of thanks event increase significantly at the first few months (Fig. 2). In the Japanese Wikipedia, the number of thanks event is still growing (2.3 times compare to January 2014) while the number of thanks was saturated or decreased in the Korean Wikipedia (0.8 times compare to January 2014.

---

[1] http://blog.wikimedia.org/2011/07/01/positive-feedback-works-for-editing-say-wikipedia-editors/
[2] http://en.wikipedia.org/wiki/Wikipedia:Notifications/Thanks
[3] http://www.mediawiki.org/wiki/WikiLove

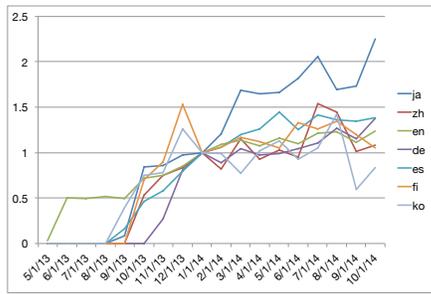

*Fig. 1: Temporal change of monthly # of thanks events (compare to January 2014)*

The population of wiki-thanks senders and recipients are overlapped because wiki-thanks senders could be recipients, and vice versa. However, the amount of overlap varies among languages. For instance, the highest overlap ratio is 47% in German and the lowest is 27% in Korean, which means Korean tend to be either sender or recipient while German do both. The difference between the number of sender and recipient is biggest in Japanese (1.5 times recipient) and smallest in Korean (1.02 time recipient). Regarding the mutual tie (a pair of editors who send wiki-thanks reciprocally.) ratio, Spanish has highest ratio and Japanese has lowest.

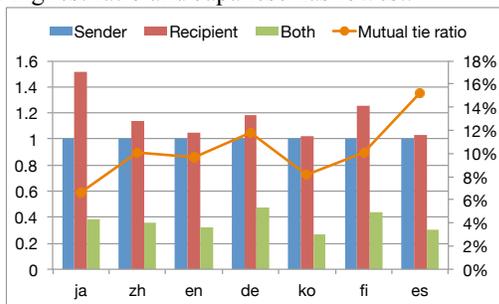

*Figure 2: Comparison among the # of sender, receiver, and sender and receiver (compare to the # of sender)*

**Demographic comparison between sender and recipient of wiki-thanks**

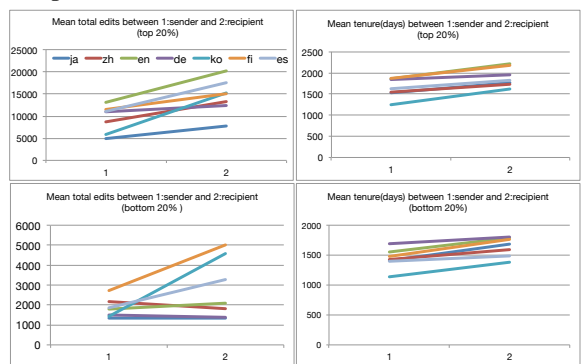

*Figure 3: Comparison total number of edits and tenure (days) between 1:senders and 2:recipients*

Figure 3 shows the demographic difference between sender and recipient from the viewpoint of Wikipedia career. The mean total edits and tenure of top 20% recipients are greater than that of top 20% of senders in all languages (significant except ko and es in total edits), which means recipients are more experienced editors. In contrast, the mean total edits of bottom 20% recipients is significantly greater than bottom 20% senders only in Korean.

**The effect of wiki-thanks feedback**

We extracted two types of editors, experienced (top 50%) and newbies (bottom 50%), based on both the total edits and the length of tenure.

*Table3: The difference between 30 days total edits before wiki-thanks and 30 days total edits after the event*

|    | Experienced to Experienced | Newbie to Experienced | $p$ | Experienced to Newbie | Newbie to Newbie | $p$ |
|----|---:|---:|---|---:|---:|---|
| ja | -8.09 | -16.97 | 0.46 | -5.64 | -5.04 | 0.63 |
| zh | -17.75 | -14.56 | 0.72 | -2.11 | -4.51 | **0.08** |
| en | -8.22 | -7.73 | 0.79 | -4.85 | -4.05 | **0.00** |
| de | -2.90 | -8.19 | **0.06** | -3.78 | -3.10 | 0.17 |
| ko | -51.54 | -51.28 | 1.00 | -6.73 | -6.96 | 0.94 |
| fi | -19.50 | -23.68 | 0.81 | -8.27 | -3.38 | 0.06 |
| es | -6.19 | -11.47 | 0.20 | -3.35 | -3.35 | 1.00 |

Overall, even though the recipients got thanks, the mean number of edits after the thanks decreased; however, the decrease ratio differs in the types of senders (Table 3). In the Chinese Wikipedia, a wiki-thanks from experienced to newbie has positive impact (reduced decrease ratio) ($p < 0.1$), which implies that a thanks from experienced editors seems to have increase motivation of newbies. In contrast, a thanks from newbie to newbie has positive impact in the English comparing to the one from experienced editors, which implies experienced users' feedback does not have change the motivation of newbies, and newbies encourages each other.

**CONLUSIONS**

The wiki-thanks feature is adopted and used in a slightly different way among languages (or cultures) in terms of mutuality, demographic, and the motivational impact although it has been only a year since wiki-thanks introduced in the Wikipedia. This study is a preliminary analysis of wiki-thanks network. In the feature work, we will look into how this social feature helps users' collaboration and the effect on creativity and performance from the viewpoint of the collaborative innovation network.